\documentclass[twocolumn,preprintnumbers,amsmath,amssymb]{revtex4}
\usepackage{tabularx,graphicx}

\usepackage{color}
\usepackage{hyperref}
\hypersetup{
    colorlinks=true,
    linkcolor=blue,
    filecolor=blue,      
    urlcolor=blue,
}

\usepackage{color}

\usepackage{ulem}   

\begin{document}

\newcommand{\beq}{\begin{equation}}
\newcommand{\eeq}{\end{equation}}
\newcommand{\beqn}{\begin{eqnarray}}
\newcommand{\eeqn}{\end{eqnarray}}
\newcommand{\bmath}{\begin{subequations}}
\newcommand{\emath}{\end{subequations}}
\newcommand{\bra}[1]{\langle #1|}
\newcommand{\ket}[1]{|#1\rangle}


\title{What  holes in superconductors  reveal about superconductivity}

\author{J. E. Hirsch}
\address{Department of Physics, University of California, San Diego,
La Jolla, CA 92093-0319}

  \begin{abstract} 
 We consider  a type I superconducting body that contains one or more holes in its interior that  undergoes a transition between normal and superconducting states in the presence of a magnetic field. We argue that unlike other thermodynamic systems that undergo  first order phase transitions the system cannot reach its equilibrium
 thermodynamic state, and that this sheds new  light on the physics of  the Meissner effect. How the  Meissner effect  occurs has not been addressed within the conventional theory
 of superconductivity, BCS. The situation considered in this paper indicates  that expulsion of magnetic field requires physical elements absent from   Hamiltonians assumed to describe superconductors within BCS theory. These physical elements are essential components of the
  alternative theory
 of hole superconductivity.
\end{abstract}
 \maketitle 
 
              \begin{figure} []
 \resizebox{7.5cm}{!}{\includegraphics[width=6cm]{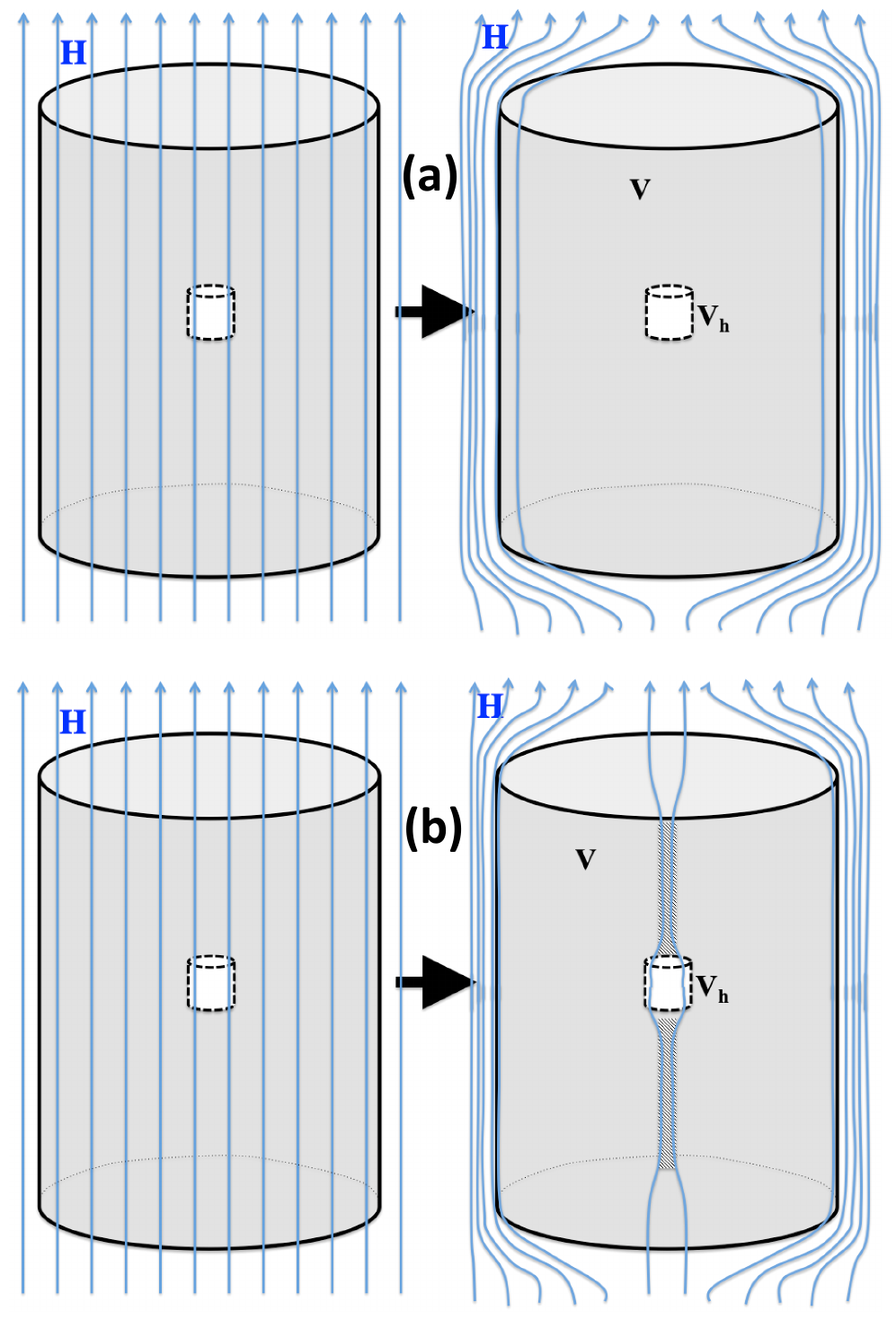}} 
 \caption { $S\rightarrow N$ transition  in the presence of a magnetic field for a sample of volume $V$ with a hole in its interior of 
 volume $V_h<<V$. (a) Magnetic field is completely excluded except within a surface layer adjacent to the outer surface of the cylinder.
 The entire material is in the superconducting state. (b) Magnetic field in the interior of the hole remains, and field lines escape through normal
 regions in the material above and below the hole. Part of the system remains in the normal phase.}
 \label{figure1}
 \end{figure}

 \section{introduction}
 It is generally believed that the transition between normal and superconducting states of a type I superconductor in the presence of a magnetic field is a first order thermodynamic phase  transition between two phases of matter         \cite{londonbook,shoenberg,tinkham,degennes},
 just like e.g. the  water-vapor or water-ice transition, that  is thermodynamically reversible and occurs with no
 change in the entropy of the universe in an ideal situation where  it  occurs infinitely slowly.
 BCS theory has not addressed the dynamic processes that take place during such a transition \cite{tinkham,degennes}. Nevertheless, it is believed that the system will find its way to reach its equilibrium lowest free energy thermodynamic state
  \cite{genforce}. Instead, within the alternative theory of hole
 superconductivity \cite{holesc,book}  it is argued that expulsion of magnetic fields requires dynamic processes that cannot occur in
 systems described by the Hamiltonians used in BCS and other such theories.
 
 To shed further light on this question we consider in this paper the situation where the material undergoing the transition contains one or more cavities (holes) in its interior.
Consider a  cylinder of volume $V$ with a small cavity in its interior of volume $V_h<<V$ as shown in Fig. 1.
 The material is a type I superconductor with critical temperature $T_c$ and thermodynamic critical field $H_c(T)$
 which monotonically increases as $T$ decreases, with $H_c(T_c)=0$. 
We ask the question:  
{\it  if the system is initially in the normal state    with a uniform magnetic field $H<H_c(T=0)$  in its interior pointing along the
 cylinder axis, will the entire system enter the superconducting state   when it is cooled 
 to a temperature $T$ such that $H<H_c(T)$}, as shown in Fig. 1 (a)? 
Note that in the analogous situation of for example  water in a bucket with
 an inclusion, when the temperature is lowered all the water would freeze into the ice phase irrespective of the presence of the inclusion \cite{flexible}. When the temperature is raised again, all the ice would melt into water, and the process would be thermodynamically reversible with no change in the entropy of the universe if it occurs infinitely slowly and involves exchange of heat with a heat reservoir at the same temperature to absorb and release the latent heat of the transition.

  According to the conventional theory of superconductivity \cite{tinkham,degennes}, BCS,  apparently the answer to this question is yes, as given for example
  in Ref. \cite{poole}. The difference in free energies of the system in  the normal state with magnetic field in its interior and superconducting state with the magnetic field excluded
  at temperature $T$  for a system of volume $V$ is \cite{tinkham}
  \beq
  F_n(T)-F_s(T)=\frac{H_c(T)^2}{8 \pi} V.
  \eeq
The lowest free energy of the system is attained when the entire system is superconducting, which requires that  there is no magnetic field 
   in the system beyond a surface layer of thickness $\lambda_L$, the London penetration depth, around the cylinder's lateral surface where a current $I$ circulates,
 that generates the magnetic field that cancels the interior field, as shown   in Fig. 1a.
From Ampere's law, assuming a cylinder of 
 very large aspect ratio $\ell/R$, with $\ell$ and $R$ the height and radius of the cylinder respectively, the current $I$  is
 \beq
 I=\frac{c}{4\pi} \ell H.
 \eeq
  BCS theory does not address the dynamic processes that take place to bring the system from the left to
  the right panel of Fig. 1 (a), it  assumes   that the system will
 find its way to reach its equilibrium thermodynamic state shown on the right panel of Fig. 1 (a), just like any other system
 that undergoes a first order phase transition would,  regardless of the presence of the inclusion.
%
 
              \begin{figure} [t]
 \resizebox{8.5cm}{!}{\includegraphics[width=6cm]{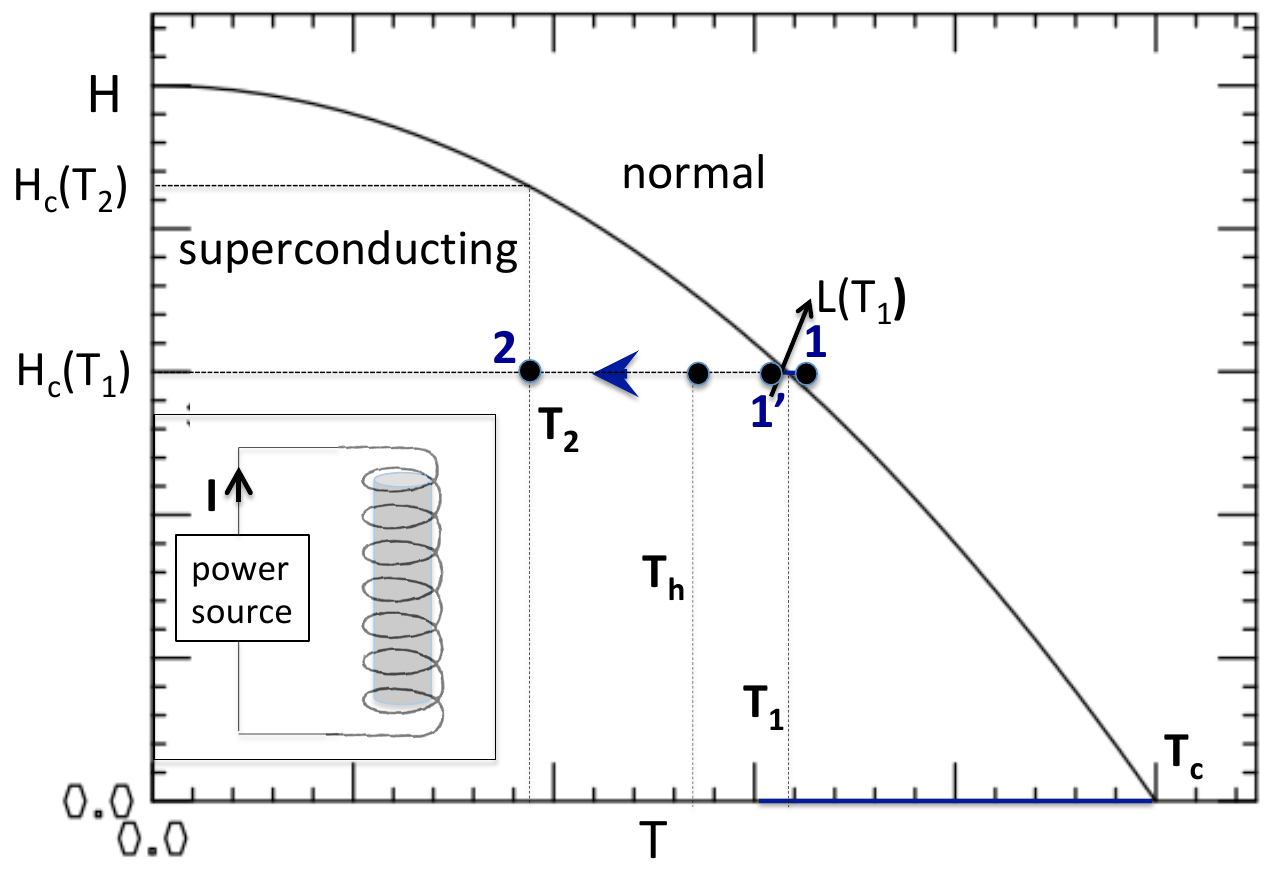}} 
 \caption {   Critical field versus temperature for a type I superconductor. The applied external field is $H=H_c(T_1)$.
 The points  {\bf 1} and {\bf 1'} 
  are at temperatures infinitesimally above and below $T_1$. $L(T_1)$ is the latent heat released by the body when it makes the transition from the
  normal to the superconducting state at temperature $T_1$. The lower left inset shows the superconducting sample in a solenoid powered by a battery that
  delivers a constant current I (Eq. (2)). $T_h$ and $T_2$ shown in the figure are discussed in the text.}
 \label{figure1}
 \end{figure}

 Instead, within the alternative theory of hole superconductivity \cite{holesc,book} the dynamic processes that lead to the expulsion of
 magnetic fields are deemed important and  are identified. Within that theory it is clear that the required dynamic
 processes are impeded by the presence of the cavity \cite{radial,scstops} so that the transition shown in Fig. 1 (a) will not take place.
 The system cooled into the superconducting state  cannot expel the entire magnetic 
 field as shown in Fig. 1 (a), but rather will end up in the higher energy state shown in Fig. 1 (b), with part of the system  remaining in the normal
 phase, irrespective of the dimensions of the hole.  Moreover, and consistent with it, the dynamic processes involved in the reverse transition to the one shown in Fig. 1 (a) cannot occur without entropy generation  even if the transition happens infinitely slowly \cite{scstops}.
 
 \section{thermodynamics}
 Fig. 2 shows the phase diagram of a type I superconductor in a magnetic field
 $H=H_c(T_1)$ generated by a solenoid connected to a power source with current $I$ of magnitude given by Eq. (1) flowing through it.
The system undergoes a reversible first order phase transition from the normal to the superconducting
 state when the temperature is lowered from slightly above $T_1$ to slightly below $T_1$, and from the superconducting to the normal state when the
 temperature is raised from slightly below $T_1$ to slightly above  $T_1$. The transitions happen infinitely slowly and no entropy is generated
 in either process \cite{londonbook,shoenberg,tinkham,degennes,reversible}.
 
 The preceding can happen for a sample with no hole. However, it cannot happen for the system with a hole in the interior. This can be
 seen as follows:
 in expelling the magnetic flux $\phi$  from its entire volume, since the Faraday electric field generated during the process is in the same direction as the current in the solenoid (counterclockwise as seen from the top), the power source 
 supplying a constant current $I$ receives energy in the amount
 \beq
 W=-\int dt I \frac{1}{c}\frac{\partial \phi}{\partial t}=\frac{H_c(T_1)^2}{4\pi }V
 \eeq 
 where $V$ is the entire volume of the cylinder (superconducting material plus empty hole). The magnetic field energy decreases by
 \beq
 W_H= \frac{H_c(T_1)^2}{8\pi }V
 \eeq
 which means that the system plus the environment have to supply energy
 \beq
 W_{needed}= W-W_H= \frac{H_c(T_1)^2}{8\pi }V
 \eeq
 to the power source, that stores it as internal energy, with no change in its entropy.
 The energy  that the system supplies  
   is, using Eq. (1)
 \beqn
 W_{sys} &=&F_n(T_1)-F_s(T_1)+T_1(S_n(T_1)-S_s(T_1)) \\ \nonumber
 &=&\frac{H_c(T_1)^2}{8\pi}(V-V_h)+L(T_1)(V-V_h)
 \eeqn
 where $S_n$ and $S_s$ are entropies in the normal and superconducting states  and  $L(T_1)$ is the latent heat of the transition per unit volume. Since the system lowers its entropy in the transition by $L(T_1)(V-V_h)/T_1$ and the total entropy of the
 universe cannot decrease, the latent heat energy has to be delivered to a heat bath, which will  increase its entropy by $L(T_1)(V-V_h)/T_1$ if it is
 at temperature $T_1$, thus
 leaving the entropy of the universe unchanged. This means that the amount of energy available to be supplied to the battery is
 \beq
 W_{avail}=\frac{H_c(T_1)^2}{8\pi}(V-V_h)
 \eeq
 smaller than what is needed, Eq. (5), by the amount
 \beq
 W_{missing}=\frac{H_c(T_1)^2}{8\pi}V_h .
 \eeq
 Thus, the process shown in Fig. 1 (a) cannot occur if the system is cooled to a temperature 
 infinitesimally below $T_1$ without violating energy conservation. 
 It is of course not possible for heat from the heat reservoir to supply the missing energy Eq. (8)
 unless the second law of thermodynamics is violated, which has been argued is possible
 in the Meissner effect \cite{nikulov} but we will assume is not.
 
 The process does become energetically possible if the system is cooled to a temperature lower than $T_1$, so that more condensation energy
 becomes available. In particular, if the system is cooled to temperature $T_h$ such that
\beq
\frac{H_c(T_h)^2}{8\pi} (V-V_h)=\frac{H_c(T_1)^2}{8\pi} V
\eeq
or to a lower temperature. We discuss this possibility in the next section.

In contrast, it is energetically possible for the system cooled to temperature infinitesimally below $T_1$ to undergo the process shown in Fig. 1 (b), where the magnetic field remains in the interior of the hole and magnetic field lines
reach  the exterior through a flux tube where the system is in the normal state.
Assume for simplicity the hole is also a cylinder, of radius $r_0$. 
The magnetic flux remaining in the sample is $\phi_h=\pi r_0^2 H$, and a cylindrical region of radius $r_0$ remains
normal above and below the hole. The energy delivered to the power source is now
\beq
W_h=\frac{H_c^2}{4\pi }(V-\pi r_0^2 \ell) .
\eeq
Half of it is accounted for by the decrease of magnetic field energy in the volume $(V-\pi r_0^2 \ell) $, and the other half
\beq
W_{h,needed}=\frac{H_c^2}{8\pi }(V-\pi r_0^2 \ell) 
\eeq
is supplied by the part of the system that condenses into the superconducting state, namely
\beq
W_{sys,h}=(\frac{H_c(T_1)^2}{8\pi}+L(T_1))(V-\pi r_0^2 \ell)
\eeq
since neither the hole nor the flux tube that stays in the normal state supply condensation   energy.
The transition shown in Fig. 1 (b) can proceed satisfying energy conservation and leaving the entropy of the universe unchanged,
and is reversible.

\section{supercooling}

As discussed in the previous section, the transition shown in Fig. 1 (a) cannot occur if the system is cooled to a temperature infinitesimally below
the transition temperature under the applied field $H$, $T_1$, with $H=H_c(T_1)$, because it would violate energy conservation. 
It is also clear that if the system is cooled to temperature infinitesimally below $T_1$ and reaches the state
shown in Fig. 1 (b), it will not evolve to the lower energy state shown in Fig. 1 (a) if the system is subsequently  cooled
to a  lower temperature $T_2$: the magnetic field lines cannot traverse the superconducting regions, so the magnetic field will remain trapped and the system will remain in the metastable
state shown in Fig. 1 (b) upon further cooling. The normal regions, which at temperature $T_1$ were single flux tubes of radius $r_0$ above and
below the hole with field
$H_c(T_1)$ in their interior, will break up into a mixed state with thinner flux tubes containing field $H_c(T_2)$ at the lower temperature $T_2$, 
preserving the same total flux $H_c(T_1)\pi r_0^2$.

The transition shown in Fig. 1 (a) could in principle occur satisfying conservation of energy if the system in the normal state
 is rapidly  supercooled  to a temperature $T_2$ lower than the temperature
$T_h$ given by Eq. (8), and subsequently undergoes the transition to the superconducting state at $T_2$. If the system is supercooled to temperature infinitesimally below $T_h$ the transition  would have to occur without any dissipation of
Joule heat to satisfy energy conservation. This can also   be ruled out on theoretical grounds: since the system is supercooled by a finite
temperature difference $(T_h-T_1)$, there will certainly be regions where the phase boundary between normal and superconducting regions
moves at a finite rate, in which case the induced
Faraday electric field will generate finite Joule heat in the normal regions and not enough energy would be available to supply to the power source. 

What if the system is supercooled to temperature $T_2<<T_h$ and then enters the superconducting state? 
From energetic arguments alone it would not be possible to  rule  out that the system could 
``find its way'' to the lowest energy state, for example by creating initially a cylindrical current that
encloses the hole as shown in Fig. 3, that would then expand to reach  the final state with
all flux excluded shown in Fig. 1 (a). 
The transition  would satisfy energy conservation,
with an amount of Joule heat \cite{jouleheat}
\beq
Q_J=(\frac{H_c(T_2)^2}{8\pi}-\frac{H_c(T_1)^2}{8\pi})(V-V_h)-\frac{H_c(T_1)^2}{8\pi} V_h
\eeq
dissipated in the process.

However, it should be noted that this Joule heat would be smaller than the Joule heat dissipated in the same process 
by a solid sample of volume $(V-V_h)$ with no hole. This seems implausible, since assuming the samples have
the same height the one with the hole would expel more flux, and it would happen at the same rate or a faster rate
 due to the presence of the hole, thus generating more rather than less Joule heat. However we are unable to rule
 this possibility out theoretically based on thermodynamic arguments.
 
 Thus, it would be of interest to conduct experiments in samples with small holes in the interior where the sample
 is rapidly supercooled to temperature well below the transition temperature $T_c(H)$  in the presence of a field $H$, and test whether 
 the system could find its way to where it is driven by its thermodynamic goal to lower its free energy as much as possible, namely
 complete flux expulsion, particularly if the hole is very small, or at least more flux expulsion than if the transition happens infinitely slowly and flux remains trapped
 as shown in Fig. 1 (b). If experiments show that this is possible, it would invalidate the conclusions of this paper.
 
               \begin{figure} [t]
 \resizebox{8.5cm}{!}{\includegraphics[width=6cm]{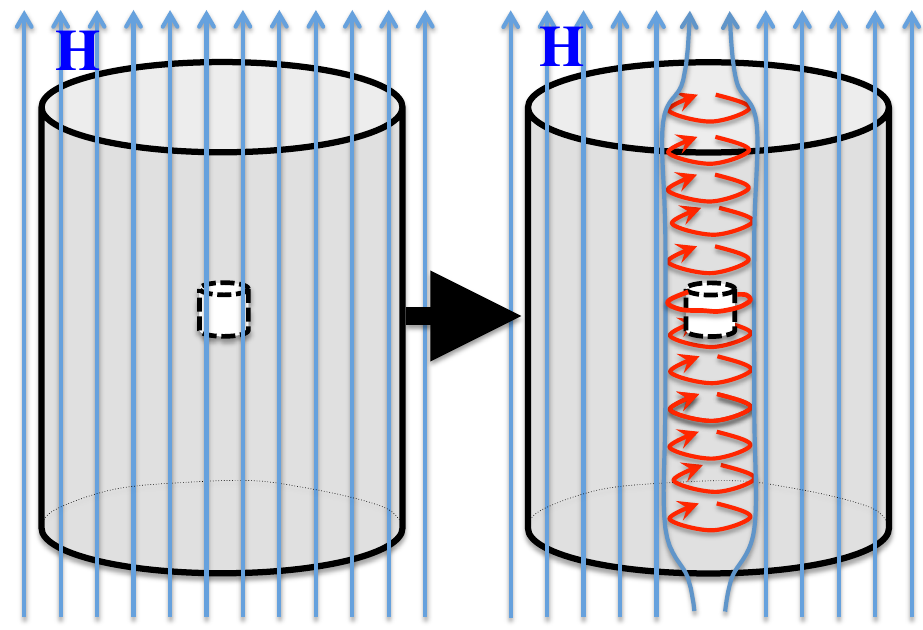}} 
 \caption { Initial stage of flux expulsion that could lead to the total flux expulsion shown in Fig. 1 (a)
 within the conventional theory.  
 The red arrows show  current flow that suppresses the field in a cylindrical region that includes the hole. }
 \label{figure1}
 \end{figure} 

\section{electromagnetic energy flow}
According to Poynting's theorem, 
\beq
\oint \vec{S}\cdot d\vec{a}=-\frac{\partial}{\partial t} \int d^3r u - \int d^3r \vec{E}\cdot \vec{J} .
\eeq
with
\bmath
\beq
\vec{S}=\frac{c}{4\pi} \vec{E}\times\vec{B}
\eeq
\beq
u=\frac{1}{8\pi} (E^2+B^2).   
\eeq
\emath
Let us apply this theorem to the surface of a cylindrical hole  of radius $r_0$ and height $h$ in the interior of the superconductor. 
The vector $d\vec{a}$ is the normal to the surface pointing outward, and the left-hand-side of Eq. (14) is the outflow of electromagnetic energy.
The first term on the right-hand-side of Eq. (14) is the decrease in electromagnetic energy inside the volume enclosed by the surface,
and the second term is minus the work done by the electric  field on charges inside the volume: if the electric field does negative work on the charges,
the outflow of electromagnetic energy increases.
The electric field on the lateral surface that is induced as the magnetic  flux $\phi$ through the hole changes is
\beq
\vec{E}(r_0)=-\frac{1}{2\pi r_0 c}\frac{\partial \phi}{\partial t}\hat{\theta}
\eeq
pointing in counterclockwise direction (as seen from the top), so the Pointing vector is parallel to the outward normal to the surface $d\vec{a}$.
Assuming  that the magnetic field at the surface of the hole stays constant at $H=H_c(T_1)$  as the magnetic field  inside the hole is expelled, the left side of Eq. (14) is
\beq
 \oint \vec{S}\cdot d\vec{a}=-\frac{1}{4\pi} h H \frac{\partial \phi}{\partial t}=-\frac{1}{4\pi} h \pi r_0^2  H \frac{\partial B}{\partial t}
\eeq
The first term on the right-hand-side of Eq. (14) is, assuming the magnetic field is uniform inside the hole
\beq
-\frac{\partial}{\partial t} \int d^3r u=-\frac{1}{8\pi} h \pi r_0^2 \frac{\partial B^2}{\partial t} 
\eeq
and Eq. (14) yields, with $V_h=\pi r_0^2 h$ the volume of the hole
\beq
-\frac{1}{4\pi}   H \frac{1}{\pi r_0^2}  \frac{\partial \phi}{\partial t}= -\frac{1}{8\pi} \frac{\partial B^2}{\partial t} -\frac{1}{ V_h}\int d^3r \vec{E}\cdot \vec{J} .
\eeq
Integrating over time and using that $B(t=0)=H$, $B(t=\infty)=0$ in the interior of the hole yields
\beq
\frac{1}{4\pi}   H^2 =\frac{1}{8\pi} H^2  - \frac{1}{V_h}\int_0^\infty dt\int d^3r \vec{E}\cdot \vec{J} .
\eeq
Eq. (20) shows that the magnetic field cannot be expelled from the interior of the hole unless the electric field in the interior of the  hole performs
negative work on charges inside the hole in the amount $-H^2/(8\pi) V_h$. Since there are no charges inside the hole, 
Eq. (20) cannot be satisfied, which means that the applied magnetic field $H$ cannot be expelled from the interior of the hole.

If instead the cylindrical volume of radius $r_0$ and height $h$  is in a region of the material without a hole, Eq. (20) will be satisfied as the material inside the hole goes from the normal to the superconducting state because a supercurrent flowing clockwise gets generated and the Faraday field $E$ pointing counterclockwise
performs negative work on the supercurrent  amounting to precisely $-H^2/(8\pi)V_h$ \cite{scstops}, rendering Eq. (20) an equality.

The above conclusion was reached assuming  that the magnetic field at the boundary of the hole remains unchanged as the field is expelled from its  interior.
If instead the magnetic field at the boundary of the hole was  to go to zero as the magnetic field is expelled, Eq. (14) would be satisfied with no contribution from
its second right-hand term, i.e. no current in the interior of the hole. This would happen if, for example, the Meissner current expelling the magnetic field
would develop at the boundary of the cylinder, so that the magnetic field would decrease uniformly inside the cylinder including the region of the hole
and its surface. This is however impossible even within the conventional theory, since in order for the current at the boundary of the cylinder to overcome the 
Faraday counter-emf requires condensation energy from the  interior, which is not available while any magnetic field is still in the interior. 
More generally we note that  Eq. (14) implies  that for the magnetic field to go from finite to zero at any point inside the volume requires that
 at some point during the process there was electric current at that point, whose carriers gained kinetic energy through 
 the condensation process and  subsequently delivered that energy to the electromagnetic field through the 
 second term
 on the right-hand side of Eq. (14).

Still, we cannot rule out that within the conventional theory there could be other processes by which the magnetic field could be expelled from the hole
 allowing the system to find its way to the thermodynamic equilibrium state with the flux completely expelled.
  For example, in the initial stages of the process a superconducting region in the region above the hole could grow to a diameter larger than the diameter of the hole and then move downward to enclose the hole, suppressing the magnetic field in its interior and its surroundings.


               \begin{figure} [t]
 \resizebox{8.5cm}{!}{\includegraphics[width=6cm]{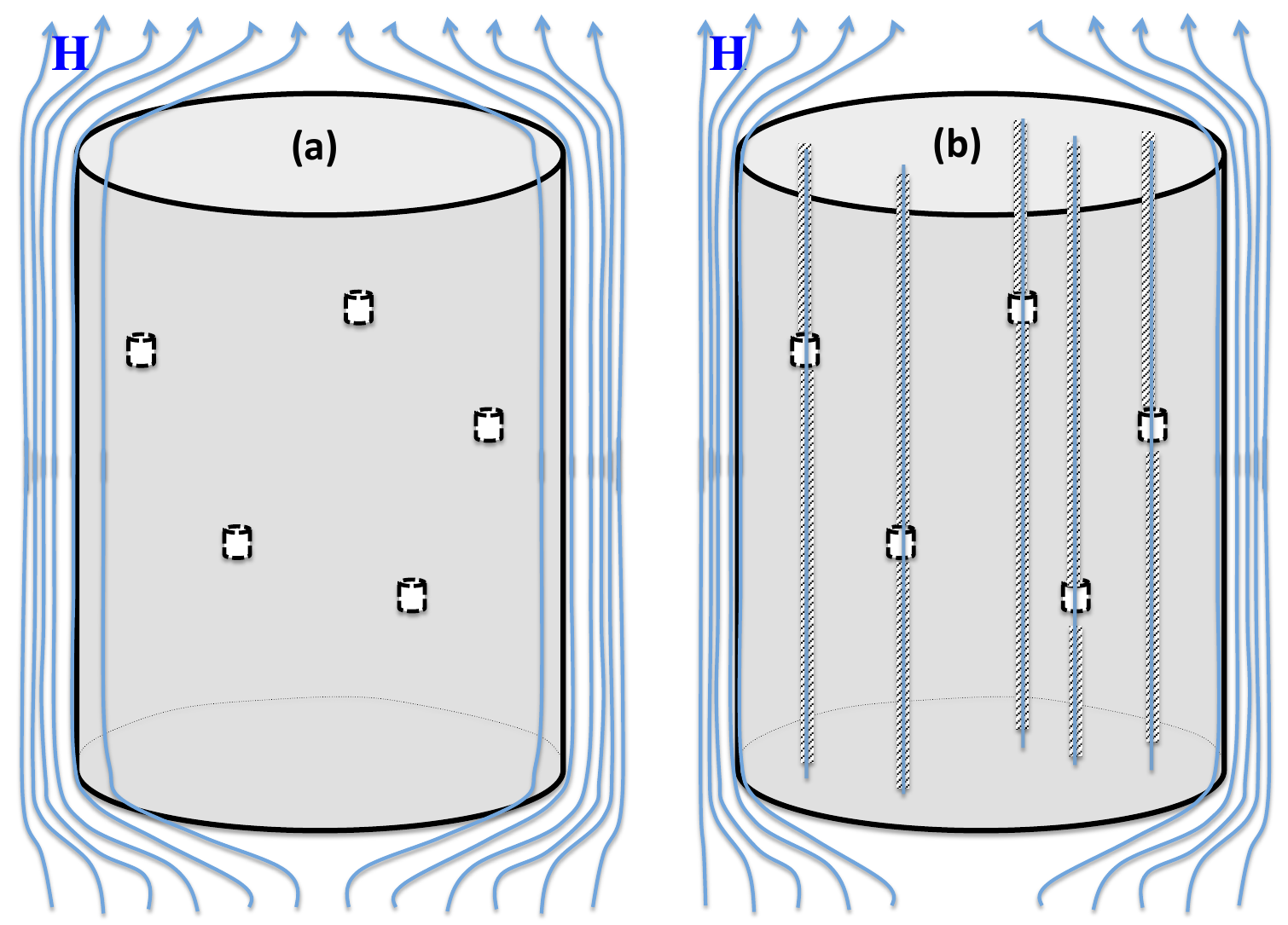}} 
 \caption {   A superconducting simply connected body with  small  holes in its interior.
 The system will reach the state (a)  with the field fully excluded if  the field is applied after the system is cooled into the superconducting state,
 and the state (b) if the normal system is cooled into the superconducting state in the presence of a magnetic field according to the theory of hole superconductivity. }
 \label{figure1}
 \end{figure} 

\section{consequences}
The considerations in the previous sections suggest that whenever there is a hole inside a solid superconducting
body, flux lines in the interior of the hole would remain when the system is cooled into the superconducting state in the presence
of a magnetic field.
Thus, the system would not find its way
to the lowest free energy final state where it is all superconducting and all the magnetic flux is excluded. Instead, a portion of the material would stay in the higher
energy normal state to allow the magnetic field lines in the hole to go out to the exterior of the cylinder: the system would trap flux wherever there is a hole,
as shown in Fig. 4, no matter how small the hole is (see Sect. VIII for further discussion on size) and no matter how
the cooling takes place.

The conventional theory of 
superconductivity does not   offer a rationale for why this  would have  to happen. Within the conventional
theory, the state of the system shown on the right panel of Fig. 1 (a) is the thermodynamically stable state
of the system under the given conditions. The phase and the amplitude of the order parameter are constant throughout
the interior, and the free energy is lowest. The cost in energy in keeping the magnetic field trapped
as shown on the right panel of  Fig. 1 (b)  is many 
times larger than the ``missing'' condensation energy of the  volume of the hole.
To understand why the system would choose to stay in this high energy state, even if it is cooled to very low temperatures, assuming experiments show that
it does,  it is necessary to 
understand what are the processes by which magnetic fields are expelled from the interior of a metal becoming superconducting.  
This understanding should explain why, unlike in any other first order phase transition, the presence of a hole
in the interior would prevent the system from reaching its thermodynamically stable state.


The discussion in the previous section indicates that magnetic field expulsion is  a $local$ process.
Magnetic field will get excluded locally from the region that becomes superconducting as it becomes superconducting.
Thus, the transition must proceed through nucleation and growth of the lower temperature phase, just as, for example, 
the water-ice transition.  However, unlike the water-ice or any other first order phase transition, here there is $momentum$ involved as well as {\it kinetic energy}. A nucleating superconducting region acquires a current circulating around its boundary
carrying momentum and kinetic energy 
that did not exist in the normal phase. This raises questions about conservation laws \cite{momentum}  that need to be answered
if one is to claim that the process is understood \cite{bcs50}, questions that do not arise in other
first order phase transitions where only potential energy is involved.

Figure 5 shows schematically what happens as the system goes superconducting
and expels the magnetic field. Superconducting domains nucleate at random points, generating current
that flows in clockwise direction as seen from the top, to nullify the magnetic field in their interior.
We assume circular domains for simplicity.
The domains  expand and
merge with other domains, with their interior currents cancelling out as they merge. When they encounter a hole,
the clockwise currents of the domains surrounding the hole create a counterclockwise current
around the hole that maintains the original magnetic field in its interior. The field in the
interior of the hole cannot be expelled because no superconducting domain can nucleate in the
interior of the hole. 

              \begin{figure} [t]
 \resizebox{8.5cm}{!}{\includegraphics[width=6cm]{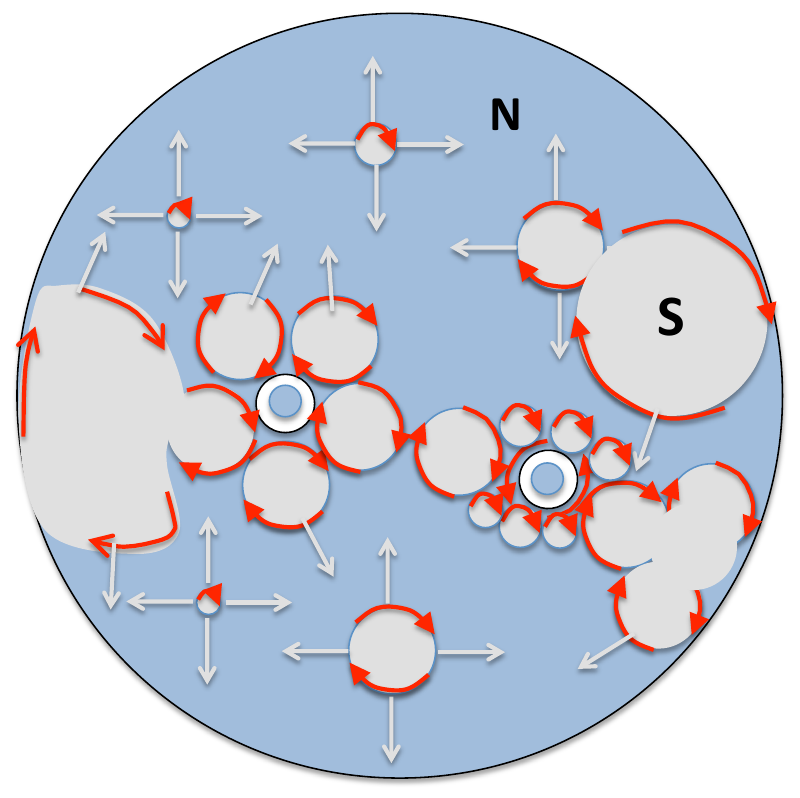}} 
 \caption {   Cross section of cylinder of Fig. 4 as seen from the top, showing normal (N) regions (blue)
 with magnetic field pointing out of the paper and superconducting (S) regions (grey) with no magnetic
 field in their interior, and two holes (white annulus with blue circle in the interior indicating trapped field).   Growing superconducting regions (grey) push out the
 magnetic field from their interior through the clockwise currents (red lines) that flow within a London
 penetration depth around
 their boundary. As domains expand they encounter other
 domains and merge with them, with the currents at the merging boundaries canceling out since they flow in
 opposite direction. 
 A superconducting domain is born and expands radially outward from a point.
 Where there are holes (white regions), no superconducting phase can nucleate to expel the magnetic
 in their interior (blue). The domains outside the holes expand until they reach the boundary of the
hole, generating a counterclockwise current around the hole that sustains the magnetic field in
 its interior.  }
 \label{figure1}
 \end{figure} 

The scenario shown in Fig. 5 follows naturally from the dynamical processes envisioned to occur at  the transition within
the theory of hole superconductivity \cite{holesc,book}, whether or not a magnetic field is present, as reviewed in the next sections.
Instead, the sole explanation of the conventional theory for the physics shown in Fig. 5 would be: the
 superconducting phase has lower free energy than the normal phase, and it can only exist without
 magnetic field in its interior. Therefore, what is shown in Fig. 5 may happen, leading to the metastable
 state shown in Fig. 4b, but more likely  the system will be driven by thermodynamics to the lowest free energy state of the entire system  
 with only a surface current and the entire magnetic field excluded as shown in Fig. 4a through other unspecified processes. 
 
 In the following we discuss what needs to happen in more detail.

\section{Maxwell pressure and Meissner pressure}
The Maxwell stress tensor in the absence of electric field is
\beq
T_{ij}=\frac{1}{8\pi}(H_iH_j-\delta_{ij} H^2)
\eeq
and in the radial direction it is given by
\beq 
T_{rr}=-\frac{1}{8 \pi}H_c^2\equiv -P_{Ma}
\eeq
This   ``Maxwell pressure'' $P_{Ma}$ resists the magnetic field expulsion at every point
inside the volume.  At a point where the condensation starts, a ``Meissner pressure''  \cite{londonbook} $P_{Me}$
needs to be generated pointing outward as the   superconducting phase nucleates and
subsequently grows, with the Meissner pressure slightly overpowering    the Maxwell pressure (if it happens reversibly). 
This is shown schematically in Fig. 6. The term ``Meissner pressure'' was coined by London \cite{londonbook}, who realized the necessity for its
existence but did not provide a physical explanation for its origin.

                \begin{figure} [t]
 \resizebox{8.5cm}{!}{\includegraphics[width=6cm]{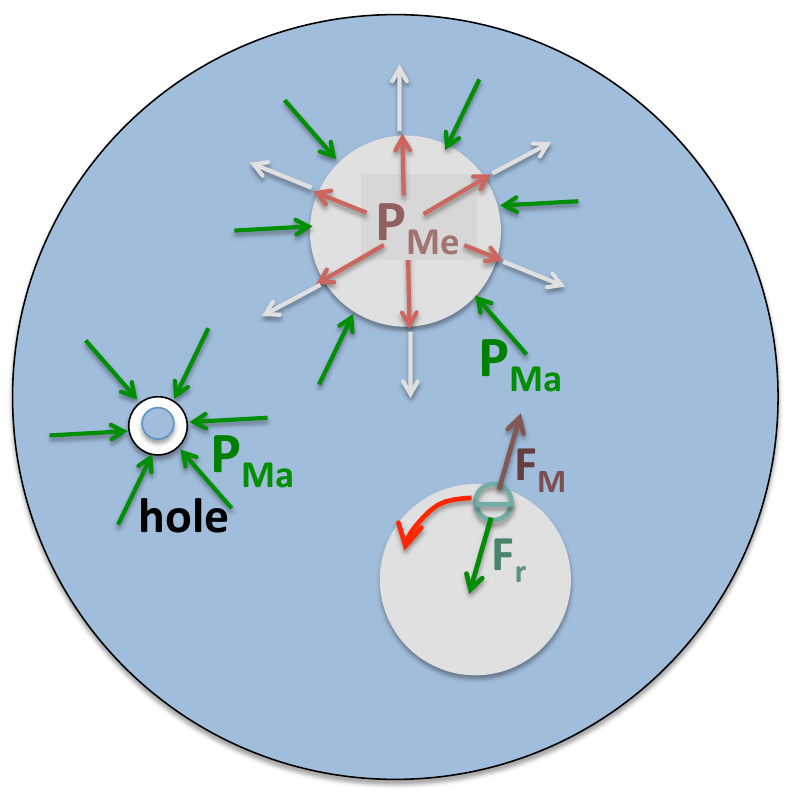}} 
 \caption {   Cross section of cylinder of Fig. 4 as seen from the top, showing how the Maxwell pressure
 $P_{Ma}$ 
 and the Meissner pressure $P_{Me}$  act. Where there is a hole (left part of the figure) there can be no Meissner pressure and the magnetic
 field cannot be expelled. The red arrow in the lower domain shows the circulation of electrons in the supercurrent near the boundary of the
 domain, subject
 to a Lorentz force $F_r$ pointing inward, balanced by a ``Meissner force'' $F_M$ pointing outward.
 }
 \label{figure1}
 \end{figure} 
 
 Where there is a hole, there can be no Meissner pressure
because there is no material inside the hole to exert such pressure, so the magnetic field cannot be expelled
from the hole nor from its immediate surroundings.
 In contrast, in an ordinary first order phase transition such as water-ice, a grain of ice would grow by water molecules
 condensing on the surface of the ice, driven by lowering of their potential energy.
 This would cause the grain of ice to grow, and there would not be an opposing ``pressure'' working against it.
 Therefore, the presence of a hole would not prevent condensation into the ice phase around the hole.

The Maxwell pressure manifests itself by the radial Lorentz force exerted by the magnetic field
on the clockwise circulating current around the boundary of a domain that nullifies the magnetic field
in its interior.
This inward force $F_r$ on an electron \cite{ondyn,koizumi}  is shown in  the domain in the lower part of Fig. 6, and is given by
\beq
\vec{F}_r=\frac{e}{c}\vec{v}\times\vec{B}
\eeq
with $\vec{v}=v_\theta(r)\hat{\theta}$ the azimuthal velocity of the electrons in the supercurrent near the boundary of the domain,
flowing in counterclockwise direction. $\vec{F}_r$ represents the transfer of momentum from the electromagnetic field
described by $T_{rr}$
to the electrons.

For the domain to grow rather than shrink, the system as it condenses needs to do work against
this radial   force  that acts  on  the electrons in the supercurrent near the surface of the domain. For a circular
domain of radius $r_0$, the velocity of the electrons is
\beq
\vec{v}_\theta(r)=-\frac{e \lambda_L}{m_e c}H_c e^{(r-r_0)/\lambda_L}\hat{\theta}
\eeq
and the magnetic field is
\beq
\vec{B}(r)=H_ce^{(r-r_0)/\lambda_L}\hat{z}
\eeq
so that the radial force is
\beq
\vec{F}_r(r)=\frac{e}{c}\vec{v}\times\vec{B}=-\frac{e^2 \lambda_L}{m_e c^2}H_c^2 e^{2(r-r_0)/\lambda_L} \hat{r}
\eeq
For the radius $r_0$ to expand by $dr_0$ the work that needs to be performed against this
inward radial force is
\beq
dW=\int d^3r n_s F_r(r) dr_0
\eeq
with $n_s$ the density of superconducting carriers. Eq. (27) yields, using that \cite{tinkham}
\beq
\lambda _L^2=m_e c^2/(4 \pi n_s e^2)\eeq, 
\beq
dW=\frac{H_c^2}{4\pi} \pi \ell r_0 d(r_0)  =    \frac{H_c^2}{8 \pi} dV
\eeq
which is the energy supplied by the system when a volume element $dV$ of the system condenses into the superconducting state. 

How does the system supply that energy as electrons condense into the superconducting state? 
Carriers condensing from the normal to the superconducting state at the phase boundary suddenly acquire kinetic energy, the kinetic energy
of the supercurrent, which is the condensation free energy Eq. (1). The kinetic energy per unit volume  at the boundary, $r=r_0$, is, from Eqs. (24) and (28),
\beq
K_s=n_s \frac{1}{2} m_e v_\theta(r_0)^2=\frac{H_c^2}{8\pi} .
\eeq
As the boundary moves further out, that kinetic energy is gradually transferred to the electromagnetic field through the second term in Eq. (14), since
the Faraday electric field points in direction opposite to the supercurrent (counterclockwise), decelerates the electrons
carrying the supercurrent  and ultimately stops them when the phase boundary
has moved out beyond a London penetration depth.
These processes are shown in Fig.  7.
That kinetic  energy Eq. (30), which came from Eq. (1), the condensation free energy, gets transferred to the power source that produces the applied magnetic field $H=H_c$
through the current $I$ given by Eq. (2) flowing counterclockwise in a solenoid surrounding the superconducting sample.

               \begin{figure} [t]
 \resizebox{8.5cm}{!}{\includegraphics[width=6cm]{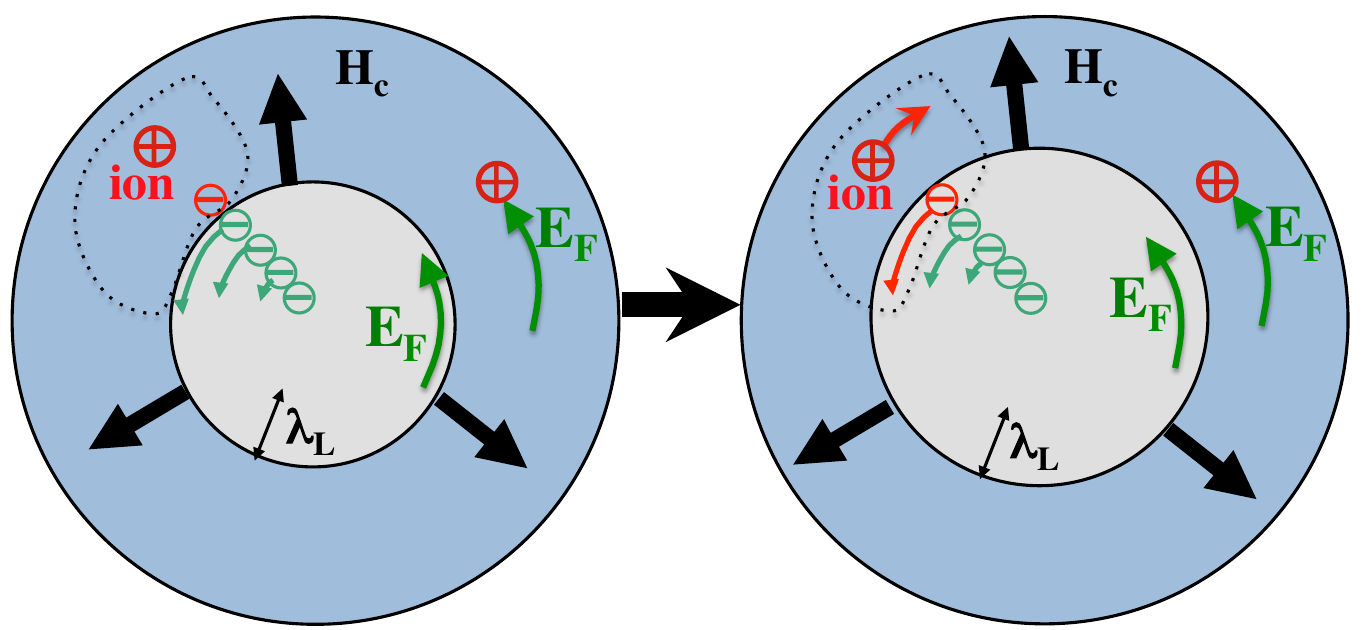}} 
 \caption { Processes involved in the expansion of a superconducting domain.
 The red electron right outside the superconducting region suddenly acquires counterclockwise momentum
 as the phase boundary crosses its position. At the same time, ions acquire clockwise momentum to ensure
 momentum conservation. The Faraday electric field acting on green electrons near the phase boundary 
 exerts clockwise force on them slowing them down as the phase boundary moves further out.
 This process transfers energy to the electromagnetic field which flows out through the lateral
 surface of the cylinder and charges the battery that supplies the current that keeps the 
 applied magnetic field constant.    }
 \label{figure1}
 \end{figure} 
 
This completely accounts for  the energetics of the flux expulsion process. But it leaves the following
key questions to be answered:

\begin{enumerate}
\item{ How does the current acquire its azimuthal momentum when a domain is born?}

\item{How does the condensation energy become  kinetic energy of the supercurrent?}

\item{ How is the azimuthal momentum of the supercurrent compensated so that angular momentum conservation is
not violated?}

\item{ How does a domain grow outward overcoming the Maxwell pressure that
wants to  shrink it?}

\item{  How does the azimuthal clockwise current  initially overcome the counterclockwise Faraday electric field
that eventually stops it?}

\item{  How does all of the above happens without energy dissipation to ensure the reversibility of the
process?}

\end {enumerate}
And of course the same questions in reverse exist and need to be answered about the reverse process, the superconductor to normal transition in a magnetic field \cite{scstops}. 
The conventional theory has not addressed any of these questions.
The  answers according to the theory of hole superconductivity \cite{holesc,book},
from which we will learn why magnetic fields from cavities cannot be expelled,  are summarized in the next section.

\section{the dynamics}
When a superconducting nucleus first forms in the presence of a magnetic field,
electrons need to form Cooper pairs $and$ acquire azimuthal velocity to oppose the existing magnetic field. In the absence
of magnetic field, electrons will pair but not acquire azimuthal velocity. Thus, {\it it is the magnetic field that imparts
azimuthal momentum to the electrons}. There is only one way that magnetic fields impart momentum to
charges, and that is the magnetic Lorentz force \cite{lorentz}. Thus, the nucleation and subsequent growth of domains
with azimuthal currents has to involve {\it radial motion} of charges, as proposed within the theory
of hole superconductivity \cite{lorentz,momentum,ondyn,onkochcherry} and earlier
independently by K. M. Koch \cite{koch} and W. H. Cherry \cite{cg}.

               \begin{figure} [t]
 \resizebox{8.5cm}{!}{\includegraphics[width=6cm]{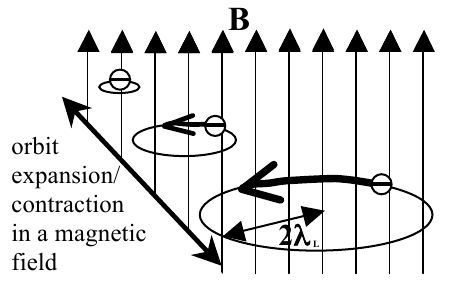}} 
 \caption {   When an electron expands or contracts its orbit
in a perpendicular magnetic field its azimuthal velocity
changes proportionally to the radius of the orbit due
to the azimuthal Lorentz force acting on the radially
outgoing or ingoing charge. }
 \label{figure1}
 \end{figure} 

According to the theory of hole superconductivity, this is accomplished by 
the enlargement of  electronic orbits, initially of microscopic size $k_F^{-1}$ ($k_F=$ Fermi momentum)
in the normal state, to mesoscopic radius $2\lambda_L$ \cite{sm} in the superconducting state, as shown schematically in Fig. 8.
The   charges   acquire azimuthal momentum  through the Lorentz force
\beq
\vec{F}_L=e\frac{\vec{v}}{c}\times \vec{H}
\eeq
acting on  charge moving radially out. That is the $only$ way to explain how the azimuthal current develops, there are no other
azimuthal forces.
The azimuthal velocity acquired by an electron expanding its orbit to radius $r$ in the presence of magnetic field
$H$ perpendicular to the orbit is \cite{copses}
\beq
v_\theta=\frac{er}{2m_e c}H
\eeq
and for $r=2\lambda_L$
\beq
v_\theta=\frac{e\lambda_L}{m_e c}H
\eeq
which is the velocity of the superconducting electrons at the boundary between superconducting and normal regions. 
Once a superconducting nucleus is born it will grow, as shown in Fig. 9:
when orbits expand negative charge moves outward, and this creates a radial electric field that draws
normal electrons right outside the domain to drift inward, expanding their radius to  $2\lambda_L$ as they enter the domain, thus pushing the boundary outward. 
The backflowing normal electrons, having negative effective mass, transmit azimuthal momentum to the
body as a whole without any scattering processes, ensuring 
momentum conservation \cite{momentum}. The processes are   shown in Fig. 10 and  discussed in quantitative detail in the references. 
The growth of a domain of radius $r_0$ at rate $\dot{r}_0$ is thus accompanied by an outward flow of 
normal $holes$ in a boundary layer of thickness $\lambda_L$ \cite{momentum}.

               \begin{figure} [t]
 \resizebox{8.5cm}{!}{\includegraphics[width=6cm]{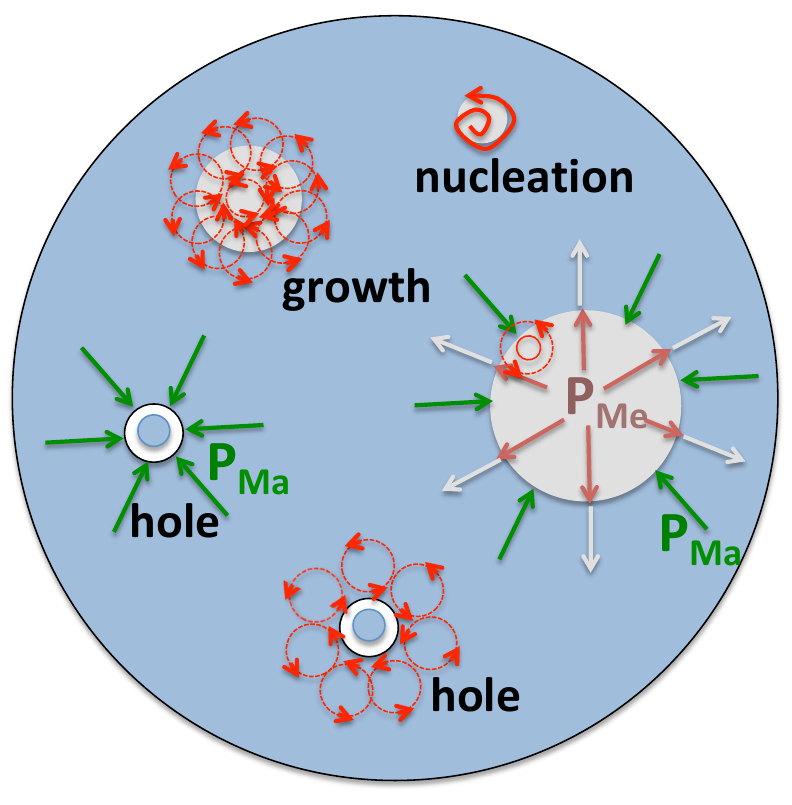}} 
 \caption {   Images showing the nucleation and growth of superconducting regions and the associated
 motion of electrons (red curves),
 the Maxwell and Meissner pressures, and how the current around a hole that keeps its magnetic field inside develops.
 }
 \label{figure1}
 \end{figure}

None of this can happen when there is a cavity (hole) inside the superconducting body. There are no electrons 
 in the interior of the cavity that can enlarge their orbits thus generating an outflow of negative charge, nor a corresponding backflow.
As a consequence, the magnetic field
in the interior of the hole cannot be expelled, consistent with the thermodynamic and electrodynamic arguments discussed in
Sects. II and IV. The electrons right outside the hole  expanding their orbits to radius $2\lambda_L$ create a net current flowing counterclockwise around the
hole, sustaining the magnetic field in its interior, as shown in the lower inset in Fig. 9.  

The above explains how the Meissner current acquires its momentum and how momentum conservation is satisfied,
questions 1 and 3 at the end of Sect. VI. The answer to question 5, why doesn't the Faraday electric field prevent 
the electrons in expanding orbits from acquiring counterclockwise azimuthal momentum, is simply that the
orbit expansion and associated outward radial motion occurs on a very fast timescale, hence the effect of the Faraday electric force imparting
clockwise momentum is negligible, as shown quantitatively in Ref. \cite{momentum}.

               \begin{figure} [t]
 \resizebox{6.5cm}{!}{\includegraphics[width=6cm]{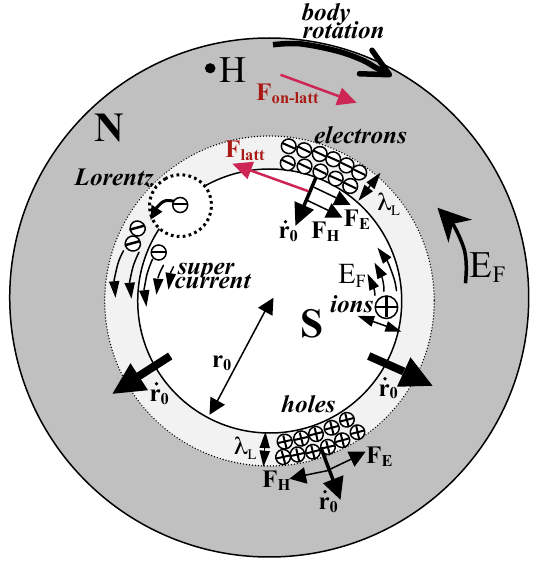}} 
 \caption { Growth of a domain, in more detail, as seen from the top.  The magnetic field points out of the paper, Meissner
current flows clockwise.$E_F$  is the Faraday electric field. The expanding orbit (dotted circle) acquires countrclockwise momentum through the Lorentz force. The backflowing carriers are shown both
as electrons moving in (upper part) or equivalently as holes moving out (lower part). 
 $F_{on-latt}$   transfers clockwise momentum to the body \cite{momentum}. }
 \label{figure1}
 \end{figure}

We still have to explain how electrons in the developing supercurrent acquire their kinetic energy and overcome
the Maxwell pressure (questions 2 and 4   in Sect. VI), since the Lorentz force does not do work on charges. The condensation energy is quantum kinetic energy \cite{kinenergy}, and the ``Meissner pressure''
is quantum pressure \cite{emf}, just like the pressure exerted by a quantum particle against the walls of a quantum well. The quantum kinetic energy of an electron confined to a region of spatial extent r is
$\sim \hbar^2/(2m_e r^2)$, hence when the  spatial extent of the wavefunction increases the kinetic energy decreases. 
In joining the condensate, condensing electrons expand their orbits from microscopic radius
$k_F^{-1}$ to mesoscopic radius $2\lambda_L$ and  this gives rise to the condensation energy lowering. 
In the presence of a magnetic field, it becomes the kinetic energy of the supercurrent at the phase boundary through
the action of the Lorentz force imparting azimuthal momentum.

%
%
The reason that normal electrons cannot expand their orbits and thereby lower their quantum kinetic energy is that it requires phase coherence, since expanded orbits are overlapping, as shown schematically in Fig. 11. The normal higher energy state without phase coherence and small nonoverlapping orbits has higher entropy and thus is preferred at temperatures above the transition temperature \cite{emf}.

                \begin{figure} [t]
 \resizebox{8.5cm}{!}{\includegraphics[width=6cm]{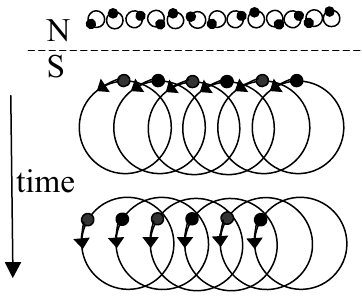}} 
 \caption { Graphic explanation for why phase coherence is required in the superconducting phase (S)  due to expansion of the orbits discussed in the text and not in the normal (N) phase. The black dots represent the position of the electron in its orbit, and its angular
 position is
 the `phase'. }
 \label{figure1}
 \end{figure} 

%
%

We have also implicitly answered question 6 of Sect. VI, how does this all happen without energy dissipation to ensure
the reversibility of the process. The answer is, there are no scattering processes involved in transferring momentum
between electrons and the ions that would cause dissipation.  The electromagnetic field plays a key role in mediating the
momentum transfer between electrons and the body as a whole. In imparting azimuthal momentum through the Lorentz force to the expanding orbit,
the electromagnetic field accumulates equal and opposite angular momentum \cite{momentum}
\beq
\vec{L}_{em}=\frac{1}{4\pi c}\vec{r}\times( \vec{E}\times \vec{H})
\eeq
since outflow of negative charge creates an outward pointing radial electric field $\vec{E}$. 
This electromagnetic angular momentum is transferred to the body as a whole by backflow of normal electrons
of negative effective mass, as discussed in the references \cite{momentum} and shown in Fig. 10. 
Thus, the momentum transfer between electrons and the body as a whole involves
no scattering processes and hence no dissipation, which is a necessary condition for the process to be reversible.
Similarly in the transition from superconductor to normal as the orbits shrink  the Lorentz force stops the azimuthal motion  without 
dissipation \cite{scstops}.

Within the conventional theory of superconductivity instead, the condensation energy is potential rather than
kinetic, and there is no physical explanation for how an attractive interaction between electrons would (a) be converted
into kinetic energy of the Cooper pairs carrying the supercurrent, nor (b) how an attractive interaction would give rise to 
an outward `pressure' that overcomes the Maxwell pressure, nor (c) how azimuthal momentum is created at
a nucleation center and increasingly so when the domain grows, nor (d) how the azimuthal momentum is compensated so that 
angular momentum conservation is not violated, nor (e) how does all that happen without energy dissipation.  Nor does it provide an intuitive explanation for how phase coherence
is established.


\section{dimensions of the hole}
We can define the initial magnetic flux going through a hole as the flux across its maximum cross-section perpendicular
to the field.
If the radius of this maximum cross-section of the hole  is of order
$\sqrt{\xi \lambda_L}$ or smaller, with $\xi$ the coherence length, the magnetic flux initially going  through
the hole with magnetic field $H=H_c(T)$ would be less than a single flux quantum $\phi_0=hc/2e$. 
According to experimental results reported by Goodman et al \cite{fluxquant}, cylindrical 
shells of $Sn$ and $In$, which are type I superconductors if sufficiently pure,  completely expel the magnetic flux from their interior when they are cooled if the initial flux
is less than half a flux quantum. 
This would suggests that the magnetic field should also be completely expelled for holes of diameter
smaller than $\sim \sqrt{\xi \lambda_L}$, in contradiction to what we expect from the considerations in this 
paper. However, it should be taken into account that the samples used in Ref. \cite{fluxquant} were
probably disordered and hence  type II superconductors, where other physics not considered in this paper would  play a role.
For type I materials we expect the considerations in this paper to apply even for dimensions
of the hole much smaller than both the coherence length and the London penetration depth, down
to atomic dimensions.
This would imply that regions of the material of dimensions much larger than a tiny  hole would remain normal
at a large cost in free energy when cooled in a field. 
It is not inconsistent with the experimental observation that the diamagnetic moment observed under
field cooling of even very pure type I superconductors is always smaller than the moment measured
under zero field cooling \cite{mend,shoen}.

  \section{discussion}
  The conventional theory of superconductivity predicts that the superconducting phase forms because
  carriers develop pairing correlations due to the electron-phonon interaction that overcomes their
  Coulomb repulsion. Theories proposed to describe so-called ``unconventional superconductors'' replace the electron-phonon 
  interaction by another `pairing glue'. In all these cases, what drives the pairing is lowering of the
  carriers' {\it potential energy}. In the presence of these pairing correlations, a quantum state involving
  a coherent superposition of such `Cooper pairs' has lower energy than the normal state, but can only
  exist in the absence of magnetic field. As a consequence, the conventional viewpoint surmises, 
  if a magnetic field is initially present the system
  will find a way to reach this lower energy state with the magnetic field excluded. 
  Of course it recognizes that to exclude the magnetic field a supercurrent needs to be generated
  to produce the field that cancels the externally applied one. Yet, surprisingly, no time nor effort has been devoted to explain the dynamical mechanism for the generation of this current, neither in the 92 years since the Meissner effect was
  discovered (with the exception of Refs. \cite{koch} and \cite{cg})  nor in the 22 years since we have been arguing that an explanation is needed  \cite{lorentz}. 
  
  In this paper we have considered the question whether or not a type I superconductor with a hole in its
  interior can reach its thermodynamic equilibrium state throughout its volume  with the magnetic field excluded
  when it is cooled in the presence
  of a magnetic field. Within the conventional viewpoint there appears to be no impediment for it,
  so it is expected that it will happen, as exposed for example in Ref.  \cite{poole}. Instead, within the alternative theory of hole superconductivity
  it cannot happen because there cannot be radial charge flow within a hole, and radial charge flow is the mechanism
  by which magnetic fields are expelled in the Meissner effect according to that theory \cite{alfven}. Thus, the two theories make
  distinctly different predictions. This could be tested experimentally under controlled conditions. 
  If the result is that the system with one or more holes in its interior even occasionally can expel the entire field, it will show  that the system somehow  finds its way to
  the lowest energy thermodynamic equilibrium state as the conventional theory predicts, hence  the dynamical processes proposed in our theory don't have to take place. If instead the result is that the system can never expel the magnetic field from an interior hole, and has to partly remain in the higher energy normal state,
  it will support the basic tenet of our alternative theory that radial charge flow is indispensable to expel
  magnetic fields \cite{alfven}, and show that energy lowering alone is not a useful criterion to explain why the Meissner effect
  takes place as the conventional theory does \cite{tinkham,degennes}.

 We have also presented thermodynamic and electrodynamic  arguments that indicate that the presence of
 a hole  impedes magnetic field expulsion. This naturally raises the question, what are the underlying microscopic
 reasons that cause this impediment? In first order phase transitions driven by lowering of potential
 energy, such as the water ice  transition, there are no microscopic processes that would impede the transition in the
 presence of a hole, and the entire system will transition to the stable low temperature thermodynamic phase, i.e. ice. There are  no obvious microscopic reasons within the conventional view of superconductivity, within which the transition is also driven by lowering of potential energy, to impede the transition.
 This is disturbing, since macroscopic constraints imposed by thermodynamics and electrodynamics  should be understandable in terms of the microscopic physics
 governing the processes. 
 Instead within the theory of hole superconductivity there are  microscopic reasons underlying the
 macroscopic constraints: the   radial
 charge flow as well as the fact that the transition is driven
 by kinetic rather than potential energy lowering, which gives rise to quantum pressure (= Meissner pressure \cite{londonbook}) that overcomes the 
 Maxwell pressure. Such an   outward pressure cannot
 exist within a hole, therefore the magnetic field in the hole and its immediate vicinity cannot be expelled.

   Within our theory electron-hole asymmetry plays a fundamental role: in the absence of charge asymmetry the dynamical processes proposed to explain field expulsion would not take place. And, it is essential that what drives superconductivity is lowering of kinetic rather than potential energy. The Hamiltonians used within the conventional theory do not
  have this physics, hence they cannot give rise to the dynamical processes that we argue are essential to explain magnetic field expulsion observed to exist in nature. 
  The microscopic Hamiltonians proposed to govern superconductivity within the theory of hole
  superconductivity do possess this physics \cite{hmlondon,color,dynhub}.
  
   In summary, we have proposed in this paper that experiments and theoretical analysis of the expulsion of magnetic fields
   from the interior of superconductors with holes can shed light on the validity or invalidity of
   different theories of superconductivity.
    \newline

   \acknowledgements
   The author is grateful to R. Prozorov, A. Frano, I. Schuller, D. Arovas and R. Dynes for discussions.


\begin{references}
\bibitem{flexible} We assume the inclusion has flexible walls, so it can accommodate a small change in its volume
which could be required depending on the route the transition takes.
\end{references}
 \end{document}